\documentclass[reprint,10pt,secnumarabic,amssymb, nobibnotes, aps, prl,groupedaddress,superscriptaddress]{revtex4-1}
\usepackage{graphicx}
\usepackage{dcolumn}
\usepackage{bm}
\usepackage{siunitx}
\usepackage[dvipsnames]{xcolor}
\usepackage{lipsum}
\usepackage{verbatim}
\usepackage{soul}
\usepackage{amsmath}
\usepackage{comment}
\usepackage{setspace}
\usepackage{pdfpages}
\usepackage{etoolbox} 
\usepackage{xurl}

\makeatletter
\patchcmd{\@outputpage@head}{\@ifx{\LS@rot\@undefined}{}{\LS@rot}}{}{}{}
\makeatother

\begin{document}
%\documentclass[reprint,secnumarabic,amssymb, nobibnotes, aps, prl,groupedaddress,superscriptaddress]{revtex4-1}
%\usepackage{graphicx}
%\usepackage{dcolumn}
%\usepackage{bm}
%\usepackage{siunitx}
%\usepackage[dvipsnames]{xcolor}
%\usepackage{lipsum}
%\usepackage{verbatim}
%\usepackage{soul}
%\usepackage{amsmath}
%\usepackage{comment}

%\newcommand{\todo}[1]{\textcolor{red}{#1}}
%\newcommand{\tocite}{\textcolor{red}{[cite]}}
%\newcommand{\gray}[1]{\textcolor{gray}{#1}}
%\newcommand{\thales}[1]{\textcolor{MidnightBlue}{#1}}

%\newcommand{\jvchange}[2]{\st{#1} \textcolor{blue}{#2}}
%\newcommand{\jvcomment}[1]{\textcolor{BurntOrange}{#1}}
%\newcommand{\jvadd}[1]{\textcolor{Red}{#1}}
%\newcommand{\jvremove}[1]{\st{#1}}

%\begin{document}
	\title{Stable positron acceleration in thin, warm hollow plasma channels}
	
	\author{T. Silva}
	\email{thales.silva@tecnico.ulisboa.pt}
	\affiliation{GoLP/Instituto de Plasmas e Fus\~ao Nuclear, Instituto Superior T\'ecnico, Universidade de Lisboa, 1049-001 Lisbon, Portugal}
	\author{L. D. Amorim}
	\affiliation{Lawrence Berkeley National Laboratory, Berkeley, California 94720, USA}
    \author{M. C. Downer}
	\affiliation{Department of Physics, The University of Texas at Austin, Austin, Texas 78712-1081, USA}
	\author{M. J. Hogan}
	\affiliation{SLAC National Accelerator Laboratory, Menlo Park, California 94025, USA}
	\author{V. Yakimenko}
	\affiliation{SLAC National Accelerator Laboratory, Menlo Park, California 94025, USA}
	\author{R. Zgadzaj}
	\affiliation{Department of Physics, The University of Texas at Austin, Austin, Texas 78712-1081, USA}
	\author{J. Vieira}
	\email{jorge.vieira@tecnico.ulisboa.pt}
	\affiliation{GoLP/Instituto de Plasmas e Fus\~ao Nuclear, Instituto Superior T\'ecnico, Universidade de Lisboa, 1049-001 Lisbon, Portugal}
	\date{\today}
	
	\begin{abstract}

    Hollow plasma channels are attractive for lepton acceleration because they provide intrinsic emittance preservation regimes. However, beam breakup instabilities dominate the dynamics. Here, we show that thin, warm hollow channels can sustain large-amplitude plasma waves ready for high-quality positron acceleration. We verify that the combination of warm electrons and thin hollow channel enables positron focusing structures. Such focusing wakefields unlock beam breakup damping mechanisms. We demonstrate that such channels emerge self-consistently during the long-term plasma dynamics in the blowout's regime aftermath, allowing for experimental demonstration.
	\end{abstract}
	
	\maketitle

Plasma-based accelerators \cite{1979Tajima} routinely provide relativistic electron and x-ray beams used in high-energy-density physics \cite{2018Wood}, nonlinear quantum electrodynamics \cite{2018Cole}, material science \cite{2016He}, and biology \cite{2015Cole}. These devices are very appealing because plasmas can sustain high-amplitude electric fields: the acceleration gradients in typical laboratory plasmas can exceed several $\SI{}{\giga\volt/\metre}$ \cite{1995Modena,2005Hogan,2012Malka,2019Gonsalves}, orders of magnitude above the breakdown threshold of most materials. Providing control over the structure of such fields in plasmas may allow an advanced generation of more compact particle accelerators and light sources.
	
Thanks to recent technological advances combined with the development of ultra-fast diagnostics \cite{2018Downer}, plasma accelerators are steadily improving the phase-space quality of the accelerated electron bunches \cite{2004Mangles,2004Geddes,2004Faure,2009Rechatin,2011Pollock,2012Weingartner}. Despite these advances, the acceleration of positron bunches in plasmas still poses long-standing fundamental questions. Positron acceleration is crucial in high energy and particle physics, where the availability of more compact linear colliders could enable new discoveries.
	
While it is possible to accomplish positron acceleration in the so-called linear regime, the corresponding acceleration gradients and efficiencies are substantially lower than when the wakefields are strongly nonlinear. Nonlinear plasma waves form when a laser or a particle bunch driver is sufficiently intense to repel nearly all plasma electrons away from the axis. In this process, plasma electrons accumulate in a thin layer that delimits a spherical region (bubble or blowout) containing only the nearly immobile background plasma ions. The resulting field structure suits electron acceleration but defocuses positrons nearly everywhere. Hence, efficient positron acceleration in nonlinear plasma waves, a crucial element for future plasma accelerator-based colliders \cite{2019Cros}, is considerably more difficult than electron acceleration.
	
Besides energy transfer from head-to-tail using long positron beams \cite{2015Corde}, controlling the wakefield structure is the way to enable simultaneous focusing and acceleration for positrons. It is possible to achieve such advanced control by using shaped drivers \cite{2014Vieira, 2015Jain} or plasmas \cite{1995Chiou, 2016Gessner,2019Diederichs}. For the latter, positron (and electron) acceleration in hollow channels is attractive because of its vanishing transverse focusing fields, which ensures emittance preservation, and enable even higher acceleration efficiencies compared to the nonlinear blowout regime \cite{1986Chen,2005Lotov,2014Litos,2018Loisch}. This tremendous potential has not yet been tapped because hollow channels are prone to beam breakup instabilities \cite{1999Schroeder,2018Lindstrom}, which pose a fundamental intrinsic limit to electron and positron energy gain. Near-hollow channels \cite{2013Schroeder} and a coaxial plasma filament \cite{2018Pukhov} can mitigate beam breakup instabilities for electrons. However, these concepts are not directly applicable to mitigate beam breakup in positron acceleration.

In this Letter, we investigate a previously unrecognized mechanism leading to the generation of thin, warm hollow plasma channels with arbitrarily small radius. These channels appear self-consistently during the long-term dynamics of nonlinear plasma waves in the blowout regime. When excited by an additional intense particle bunch driver, the resulting channel provides a wakefield structure which can stably accelerate positron bunches to ultra-relativistic energies. Because of their finite temperature, plasma electrons can accumulate at the center of the thin hollow channel, providing nearly linear focusing forces during acceleration. This wakefield structure also unlocks beam breakup suppression mechanisms reminiscent of those operating for electrons in the nonlinear blowout regime, but that have been previously inaccessible in hollow channels. We illustrate our findings with theory and three-dimensional particle-in-cell (PIC) simulations using the OSIRIS framework \cite{2002Fonseca,2013Fonseca}.
	
\begin{figure}[t]
		\centering
		\includegraphics[width=.99\linewidth]{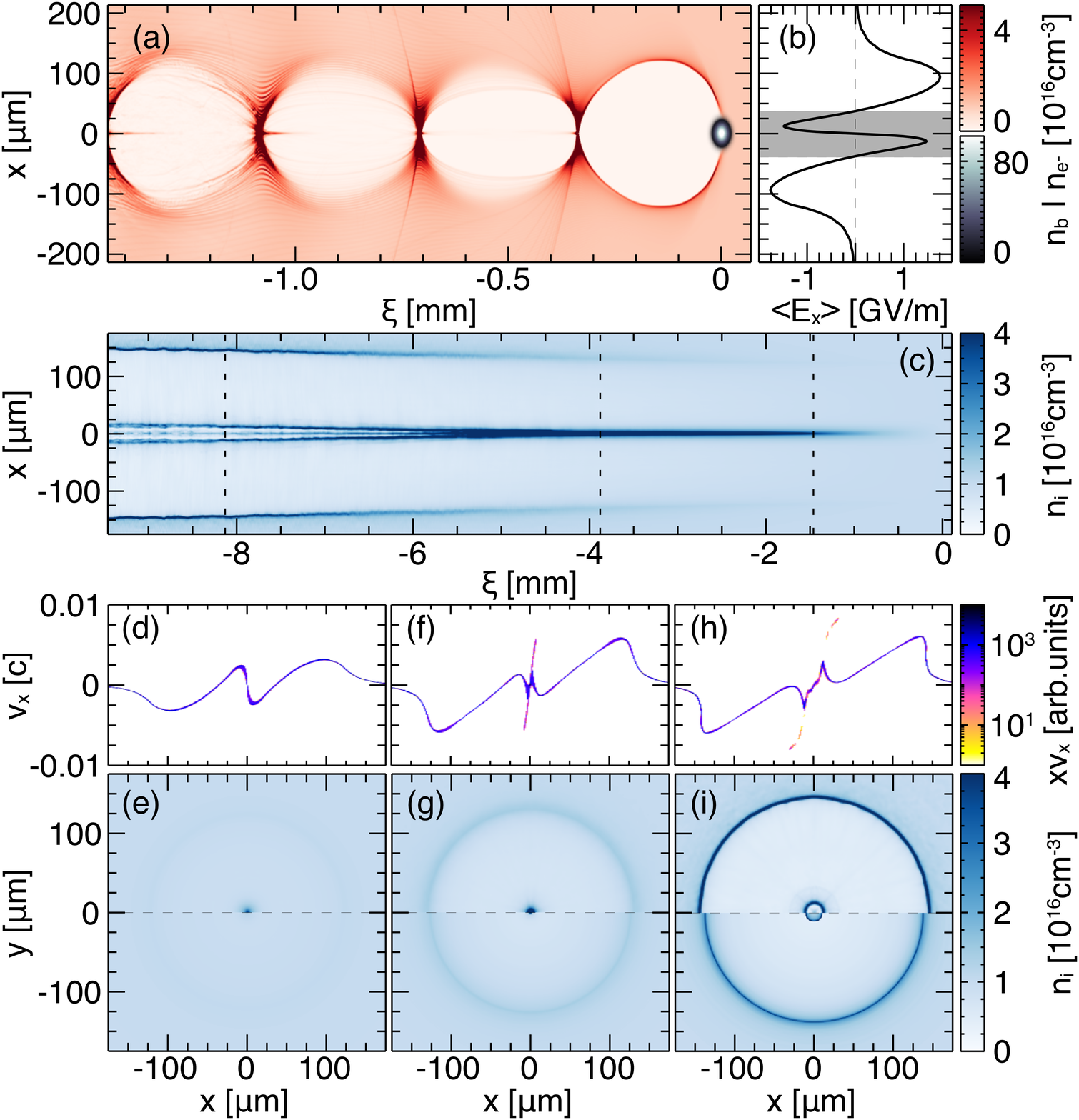}
		\caption{(a) Electron density and driver beam density. (b) Longitudinal average of the transverse electric field over the region shown in panel (a). (c) Longitudinal ion density over \SI{9}{\milli\meter} behind the driver. The dashed lines represent the position shown in panels (e), (g), and (i). (d) Ion phase-space and (e) density at $\xi \approx \SI{-1.5}{\milli\meter}$ behind the driver. The upper half of panel (e) are PIC simulation results and the lower half the semi-empirical model [Eq. \eqref{eq:1}]. Analogously, panels (f-g) and (h-i) display results at $\xi \approx \SI{-4}{\milli\meter}$ and $\xi \approx \SI{-8}{\milli\meter}$, respectively.}
		\label{fig:ion_dyn}
	\end{figure}

Figure~\ref{fig:ion_dyn} shows numerical simulation results that illustrate the onset of hollow channel formation. We consider the dynamics of a 10 GeV electron bunch [\SI{1}{\%} root-mean-square (RMS) energy spread] propagating through a preformed, uniform density $n = \SI{1e16}{\per\cubic\centi\meter}$ Hydrogen plasma. The total bunch charge is \SI{3}{\nano\coulomb}, being characterised by a bi-Gaussian density profile with an equal longitudinal and transverse size of \SI{10}{\micro\meter}. The corresponding bunch peak density is 120 times higher than the background plasma density, exciting strongly nonlinear plasma wakes in the blowout regime. The bunch has \SI{186}{\micro m} transverse emittance, which matches the beam to the blowout focusing force \cite{2002Clayton}. We find similar bunch parameters in several particle accelerators laboratories \cite{2014Tavares,2016Aschikhin,2019Yakimenko}. Simulations use a custom-built electromagnetic field solver to mitigate the numerical Cherenkov instability \cite{2017Li,2021Li}. The simulation grid has cubic cells $\SI{1}{\micro \meter}$ long; the beam, plasma electrons, and ions start with 512, 8, and 8 particles-per-cell, respectively. Figure \ref{fig:ion_dyn}(a) displays the density $n_e$ of the first few electron plasma waves in the blowout regime. Here, the variable $\xi = z - ct$ measures the distance to the bunch center, with $z$ being the longitudinal position, $t$ the time, and $c$ is the speed of light in vacuum; $x$ and $y$ are the transverse coordinates.
	
	The motion of background plasma ions plays a central role in the formation of the thin, warm hollow plasma channel. The time-averaged radial electric fields in plasma fully define the long-term ion dynamics \cite{2012Vieira,2017Sahai,2019Gilljohann,2020Zgadzaj}. Figure \ref{fig:ion_dyn}(b) provides a typical example of those fields. It shows that the average radial fields attract the ions close to the axis [gray region of Fig. \ref{fig:ion_dyn}(b)] towards $r=0$, to neutralize the excess blown-out sheath electrons that accumulate at the back of each bucket. The ion focusing region represents $1/4$ of the blowout radius, which corresponds to $\SI{25}{\micro m}$ for the specific parameters of Fig.~\ref{fig:ion_dyn}. The thin and warm hollow channel forms because of this ion focusing field region. It appears as the narrow hollow structure near the axis in Fig.~\ref{fig:ion_dyn}(c), which shows the spatial evolution of the ion density up to \SI{9}{\milli\meter} behind the driver. The time-averaged fields defocus ions sitting at larger radii and up to $\SI{200}{\micro m}$. These defocused ions accumulate at a larger radius and form the wider hollow structure in Fig.~\ref{fig:ion_dyn}(c). While the wider structure was predicted before \cite{2017Sahai}, the thin channel was neglected; we found it fundamental to stabilize positron acceleration and relax time-delay tolerances between driver and witness beam.
	
	%We now describe the ion phase-space dynamics close to the axis, as it fully defines the structure of the thin hollow plasma channel. 
	Figure~\ref{fig:ion_dyn}(d) represents the early time ion phase-space at the position of the rightmost dashed line in Fig.~\ref{fig:ion_dyn}(c). The overall phase-space structure in Fig.~\ref{fig:ion_dyn}(d) mimics the average radial field profile in Fig.~\ref{fig:ion_dyn}(b), thus confirming that the time-average radial wakefield sets the ion dynamics.  The accumulation of ions close to the axis, a result of the corresponding focusing electric field, leads to the generation of a dense ion filament, shown in the upper half of Fig.~\ref{fig:ion_dyn}(e). Weakly nonlinear plasma waves can also generate ion filaments \cite{2003Gorbunov,2012Vieira,2018Spitsyn}, thus widening the range of conditions where similar phenomena occur in experiments.
	
    An electrostatic shock \cite{2003Kaplan,2005Peano} forms when the fastest inward moving ions cross the axis. Figure~\ref{fig:ion_dyn}(f) shows signatures of this shock in the ion phase-space, whereas the upper half of Fig.~\ref{fig:ion_dyn}(g) shows the corresponding ion density profile at the position of the central dashed line in Fig.~\ref{fig:ion_dyn}(c). The electrostatic shock structure accelerates a fraction of inward moving ions to nearly twice the shock velocity, up to $\SI{0.01}{}c$. Besides, the ion motion leading to the shock also induces wavebreaking \cite{2012Vieira}, which heats plasma electrons and suppresses radial (and longitudinal) wakefields. In the absence of radial electric wakefield components, the shock front expands at a nearly constant velocity. This is consistent with Fig. \ref{fig:ion_dyn}(h), which illustrates the shock front expansion in the ion phase-space at the position of the left dashed line in Fig.~\ref{fig:ion_dyn}(c). The ions at the expanding shock front form a thin, near-hollow channel structure.
    %, suitable for stable positron acceleration.
    Figure~\ref{fig:ion_dyn}(i) shows the thin, near-hollow channel ion density transverse profile. Figure~\ref{fig:ion_dyn}(i) also shows the accumulation of ions at larger radii, close to the blowout radius, at around $r=\SI{150}{\micro m}$. An idealized version of the plasma density profile shown in Fig.~\ref{fig:ion_dyn}(h), but without plasma inside both channels, provides a solution to avoid radiation losses in electron acceleration towards \SI{}{\tera eV} energies \cite{2020Farmer}. Hence, the structures created self-consistently during the long-term plasma evolution can also be beneficial for electron acceleration.
	
	These observations suggest a simple semi-empirical model to predict the ion dynamics, and the formation of near-hollow channels observed in the simulations which is
	\begin{equation}
	\ddot{x} + \frac{Ze}{m_i}\left[\left<E_x^{}\right>\left(x\right)\right]\left(1-\frac{t}{t_{wb}}\right) \Theta\left(t\right)\Theta\left(t_{wb}-t\right) = 0,
	\label{eq:1}
	\end{equation}
	where $e$ is the elementary charge, $Ze$ and $m_i$ are the ion charge and mass, $t_{wb}$ is the wavebreaking time, and $\Theta(t)$ is the step function. Because a predictive theory for the time-average fields in the blowout regime is not available, we estimate $\left<E_x\right>$ directly from the simulation shown in Fig. \ref{fig:ion_dyn}(b). 
	
    The model given by Eq.~\eqref{eq:1} assumes that the electric field intensity decreases linearly with time until wavebreaking occurs at $t = t_{wb}$. The shock formation time provides a figure for $t_{wb}$. The shock forms when the fastest inward moving ions, initially at $x=x_0$, reach the axis. Using Eq.~\eqref{eq:1}, this occurs after $\Delta t=\sqrt{3 m_i x_0/Ze\left<E_x\right>} \equiv t_{wb}$. For the specific parameters of Fig.~\ref{fig:ion_dyn}(b), $x_0 \simeq \SI{10}{\micro m}$, $Ze = \SI{1.6e-19}{\coulomb}$, $m_i = \SI{1.7e-27}{\kilo\gram}$, and $\left<E_x\right>\simeq\SI{1}{\giga V \per\metre}$, this gives $t_{wb} \simeq \SI{17}{\pico \second}$ or $c t_{wb} = \SI{5.1}{\milli\meter}$, which is close to the simulation result [cf. Fig.~\ref{fig:ion_dyn}(c)]. The near-hollow channel forms once the fastest ions travel from $x=0$ to $x=-x_0$, which takes $\Delta t \simeq (2/3) t_{wb}$. Thus, a near-hollow channel appears after $t_{hol} \simeq (5/3) t_{wb}$. This corresponds to $t_{hol}\simeq \SI{28.05}{ps}$ or $ct_{hol} \simeq \SI{8.5}{mm}$ for our example, which is close to simulation results. As the hollow channel formation is connected to the background ion motion, the channel density profile does not change quickly, providing a high-tolerance temporal-delay range for injecting a second beam to drive wakefields in the channel. In our example, Fig. \ref{fig:ion_dyn}(i) profile is similar for $\Delta \xi \approx \SI{1}{\milli\meter}$ around $\xi \approx \SI{-8}{\milli\meter}$; the tolerance would be even higher using higher $Z$ gases \cite{SM}.
	
	We used Eq. \eqref{eq:1} to push uniformly distributed test ions, assuming a prescribed time-averaged electric field profile corresponding to Fig.~\ref{fig:ion_dyn}(b). The semi-empirical model recovers the main features of the hollow channel described above and seen in the PIC simulations. The bottom half of panels (e), (g), and (i) in Fig. 1 illustrate the predictions of the semi-empirical model, and are directly comparable with PIC simulation results (upper half of the same plots). Quantitative differences on the thin hollow channel structure are due to the steep electric field profile near the axis [see Fig. \ref{fig:ion_dyn}(b)], which make the averaged description less accurate. Furthermore, the model does not include the physics of the collisionless shock as it considers test particles.
	
    The near-hollow channel electrons are warm, with temperatures varying from $\SI{2}{\kilo e\volt}$ to $\SI{9}{\kilo e\volt}$. This distinguishing feature enables high-quality positron acceleration in the nonlinear blowout regime as long as the hollow channel radius remains sufficiently small. In a warm electron plasma, the thin electron layer that surrounds the blowout region spreads over a larger volume compared to a cold plasma. This reduces the maximum electron density and the strength of electron defocusing fields at the back of each bucket. The ions at the hollow channel walls may therefore attract and trap some of these electrons inside the near-hollow structure. The hollow channel radius controls the effectiveness of this capture process: hollow channels with smaller radii can trap more electrons because the ion density and ensuing electrostatic fields are correspondingly higher. We define thin, warm channels as the ones in which the radius and temperature are sufficiently small and high, respectively, to generate a positron focusing field structure due to an excess of plasma electrons inside the channel. We verified with PIC simulations that the electron temperature and thin channel are both essential to generate the positron focusing structure. Simulations with thin, warm perfectly-hollow channels resulted in similar results to the self-consistent, near-hollow case.

%	Figure~\ref{fig:positron_ev} shows the results of three-dimensional OSIRIS PIC simulations that demonstrate stable, high-quality positron acceleration in warm, thin hollow channels. An electron bunch excites the initial nonlinear wakefield structure that creates the channel (as in Fig.~\ref{fig:ion_dyn}). Then, a second electron bunch excites a nonlinear wakefield in the hollow channel. We adjusted the time-delay between drivers such that the hollow channel radius is slightly larger than the positron bunch transverse size. This is a necessary condition to take full advantage of the positron focusing capabilities. The background plasma density, first bunch peak density, $m_i$, and $Z$ are free parameters that dictate $t_{hol}$, thus enabling experimental control to adjust the time-delay between drivers. Positron focusing and acceleration wakefields appear after the first plasma bucket in the channel.

	To demonstrate stable, high-quality positron acceleration in thin, warm hollow channels, we relied on a set of reduced simulations that used as input parameters the self-consistent near-hollow channel density and corresponding electron spatial temperature distribution at $\xi\approx-\SI{8}{\milli\metre}$ [see Fig. \ref{fig:ion_dyn}(i)]. This approach relaxes computational requirements, isolates all essential features of the scheme, and fully recovers the results of larger-scale simulations that include the long-term plasma dynamics when the channel driver is matched to the wakefield focusing forces.
	%We adjusted the driver charge in these simulations to ensure that the hollow channel traps blown-out electrons into regions of positron accelerating fields. 
	%In addition, 
	To stabilize wakefield excitation and further ensure high-quality positron acceleration, we also matched the driver to the focusing structure in the blowout regime \cite{2002Clayton}.
	\begin{figure}[t]
		\centering
		\includegraphics[width=.99\linewidth]{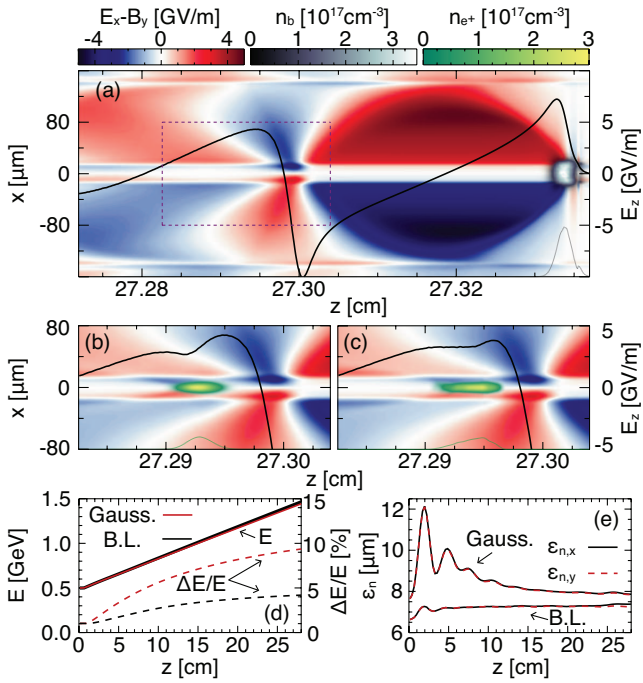}
		\caption{(a) Transverse ($E_x-B_y$) and accelerating ($E_z$, solid line) fields driven in the channel by the driver bunch [in gray]. Panels (b) and (c) are the dashed region in (a) in the presence of a Gaussian (Gauss.) and beam loading (B.L.) optimized witness positron beam, respectively. (d) Average beam energy $E$ (solid line) and RMS energy spread $\Delta E/E$ (dashed line) for the Gauss. (red) and B.L. (black) simulations. (e) Transverse emittance evolution for both examples. Panels (d-e) are functions of the propagation distance.}
		\label{fig:positron_ev}
	\end{figure}
	
    The driver in the reduced simulations is identical to the near-hollow channel driver, except that it contains \SI{1.5}{\nano C} instead of \SI{3}{\nano C}. Figure \ref{fig:positron_ev}(a) shows the driver beam density and the electromagnetic fields driven by the beam in the channel (the line is the accelerating field at $x=y=0$). It shows an extended region ($\approx\SI{150}{\micro\meter}$ long) for positron focusing and acceleration on-axis after the first plasma wave. The focusing forces vary along the longitudinal direction in the channel, thus enabling one of the mechanisms to suppress beam-breakup \cite{2017Lehe}. 
    
    We performed two reference simulations accelerating positrons; in both, the witness beam starts with \SI{100}{\pico C}, \SI{500}{\mega eV}, \SI{1}{\%} RMS slice energy spread. The positron beam is injected in the second plasma wave, around $\SI{370}{\micro\meter}$ behind the driver. Simulation parameters are the same as specified previously, except the driver and the positron beam start with 64 and 216 particles-per-cell.
    
    %JV
    The first simulation is not optimized for beam loading or emittance preservation. The witness bunch has a bi-Gaussian spatial profile with $\SI{10}{\micro\meter}$ longitudinal and  $\SI{5}{\micro\meter}$ transverse size, and a normalized emittance of $\SI{7.8}{\micro\meter}$. Figure \ref{fig:positron_ev}(b) shows the region delimited by the dashed box in Fig. \ref{fig:positron_ev}(a) in the presence of the witness bunch (in green). The positron bunch accelerates with a nearly constant accelerating gradient over the \SI{27}{\centi\meter} without beam breakup. The accelerating gradient is $\SI{3.5}{\giga eV \per \metre}$ [see Fig. \ref{fig:positron_ev}(d)], consistent with other hollow channel acceleration results \cite{2018Pukhov}. Because beam-loading is not optimal, the accelerating field varies along the beam [see Fig. \ref{fig:positron_ev}(b)]. This leads to energy spread growth [Fig. \ref{fig:positron_ev}(d)]. Still, the relative energy spread remains below $10\%$. The beam performs several betatron oscillations as it accelerates. In these oscillations, some positrons can reach regions of defocusing fields, leading to a $10\%$ reduction of the total charge at the end of the acceleration. Furthermore, these oscillations also lead to emittance variations [Fig. \ref{fig:positron_ev}(e)]. Interestingly, as a result of the dynamics of some of the bunch positrons, the final emittance is close to its initial value. Some positrons can first escape the channel as the bunch undergoes betatron oscillations, reaching the focusing region located at $|x|\approx \SI{80}{\micro\meter}$ in Fig. \ref{fig:positron_ev}(b). As some of those positrons return to the channel, they cross through a defocusing region, reducing the transverse momentum and the emittance.
    
    %JV: is Fig. \ref{fig:positron_ev}(d) ever mentioned? 
    The second example [Fig. \ref{fig:positron_ev}(c)] displays a beam-loading optimized case with near matched emittance.  The beam transverse profile is a flat-top distribution with \SI{7.5}{\micro\metre} radius and the beam starts with a normalized emittance of $\SI{6.5}{\micro\meter}$. The longitudinal current profile rises linearly in \SI{16}{\micro\metre} and falls linearly in \SI{46}{\micro\metre}. This mimics the beam-loading conditions in the blowout regime for electron acceleration \cite{2008Tzoufras}. Despite the remarkable similarity on the required longitudinal bunch current, the beam loading physics is not the same as in Ref. \cite{2008Tzoufras}. Here, higher currents at the head of the positron bunch can screen accelerating fields at those locations by attracting plasma electrons, thus flattening the longitudinal electric field structure. Similar profiles were also predicted for other positron acceleration schemes \cite{2020Diederichs}. Figure \ref{fig:positron_ev}(d) shows a similar energy gain rate as for the Gaussian beam, but with a smaller energy spread increase. The beam is closer to a matched condition, which minimizes betatron oscillations and the projected emittance growth in Fig. \ref{fig:positron_ev}(e). More than \SI{99}{\%} of the initial charge remains in the channel after \SI{27}{\centi\meter} propagation.
    
    	\begin{figure}[t]
		\centering
		\includegraphics[width=.99\linewidth]{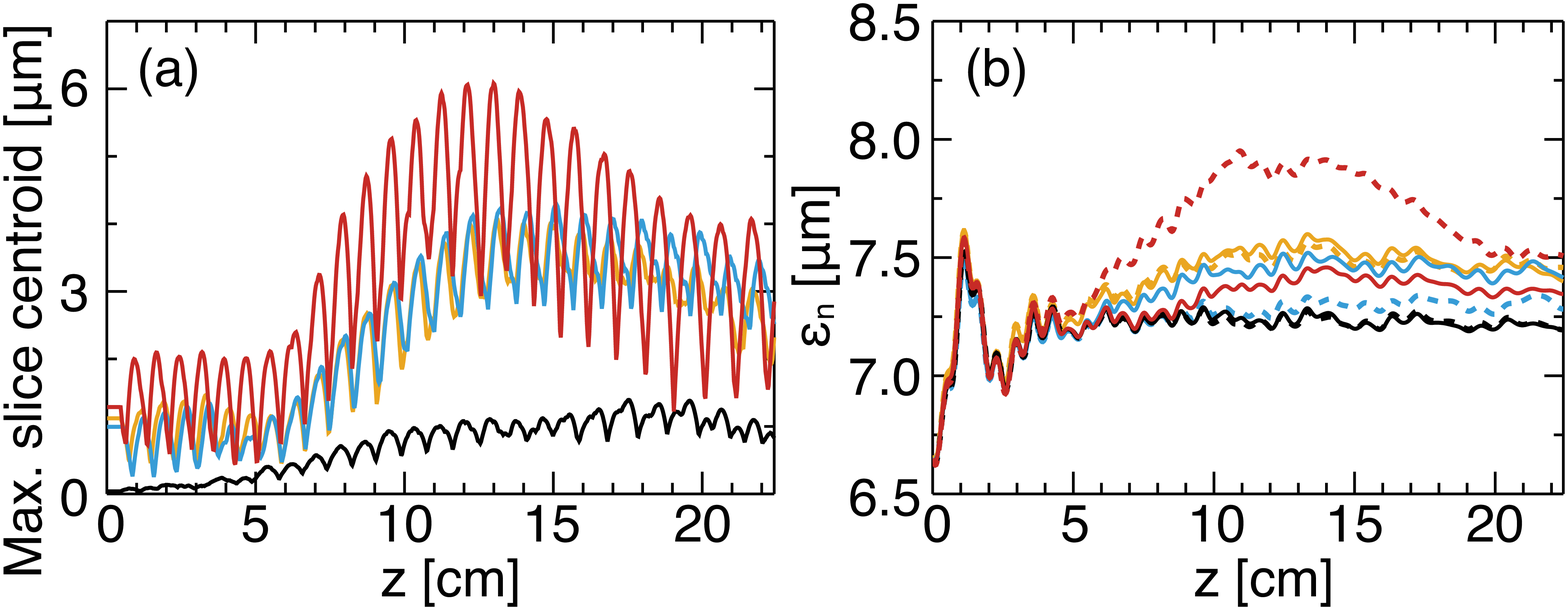}
		\caption{(a) Positron bunch maximum slice centroid evolution against propagation distance for different levels of initial centroid displacement (different colors). (b) Transverse emittance evolution for the same examples; $\varepsilon_{n,x}$ and $\varepsilon_{n,y}$ are the solid and dashed lines, respectively.}
		\label{fig:hosing}
	\end{figure}
    Thin, warm hollow plasma channels provide access to beam break-up instability suppression mechanisms, akin to BNS damping in conventional accelerators \cite{1983Balakin}. Beam breakup suppression results from the positron focusing field structure provided by plasma electrons trapped within the thin hollow channel. To show hosing instability suppression and damping, we performed an additional set of simulations identical to that in Fig. \ref{fig:positron_ev}(c) except for the initial displacement of the bunch centroid, which controls the initial seed for the hosing instability. The lines in Fig. \ref{fig:hosing}(a) illustrate the corresponding evolution of the maximum slice centroid displacement as function of the propagation distance. All simulations show hosing instability saturation and damping, with more than \SI{98}{\%} of the initial charge remaining in the bunch after $z > \SI{20}{cm}$. Figure \ref{fig:hosing}(b) shows the emittance evolution for the same examples, showing that the beam quality is not compromised; we further verified stable acceleration for different parameters presented in the supplemental material \cite{SM}. The combined action of head-to-tail variations of the focusing forces \cite{2017Lehe} and energy spread \cite{2014VieiraHosing, 2017Mehrling} activated hosing suppression mechanisms
	
    We have shown that thin, warm electron hollow channels are a previously unexplored configuration that enables stable, high-quality positron acceleration. We have established that such channels appear self-consistently during the long-term plasma dynamics in the aftermath of strongly nonlinear plasma waves in the blowout regime. While we have considered electron bunch driven wakes, similar structures may also emerge in the wake of intense laser pulses, which are common in many laboratories. Finally, besides positron acceleration, the long-term ion dynamics leading to warm electron near-hollow channels also provide the means to realize high-quality electron acceleration beyond the energy frontier \cite{2018Pukhov,2020Farmer}.
	\begin{acknowledgments}
	 We gratefully acknowledge computing time on the GCS Supercomputers JUWELS and SuperMUC (Germany); and PRACE for awarding us access to MareNostrum at BSC (Spain) and Piz Daint at CSCS (Switzerland). T.S. and J.V. acknowledge support from EU Horizon 2020 grants No. 653782 (EUPRAXIA) and 730871 (ARIES) and FCT (Portugal) grants PTDC/FIS-PLA/2940/2014, SFRH/IF/01635/2015, and UID/FIS/50010/2023. M.C.D. and R.Z. acknowledge support from NSF grant PHY-2010435 and DOE grant DE-SC0011617. M.J.H. and V.Y. acknowledge support from DOE contract DE-AC02-76SF00515.
	\end{acknowledgments}
\providecommand{\noopsort}[1]{}\providecommand{\singleletter}[1]{#1}%
%

%\bibliography{bibliography}

%\end{document}

%\clearpage
%\setcounter{figure}{0}

%\newpage
%\onecolumngrid
%{\setstretch{1.25}
%\setcounter{figure}{0}
%\renewcommand{\thefigure}{\Roman{figure}}
%\input{ref_1}
%\clearpage
%\setcounter{figure}{0}
%\input{ref_2}
%\clearpage
%\input{list_of_changes}
%}
\end{document}